\documentclass[twocolumn]{aastex631}
\usepackage{hyperref}
\usepackage{subfigure}

\newcommand{\median}[1]{\text{median}(#1)}
\newcommand{\mad}[1]{\text{MAD}(#1)}

\begin{document}

\title{COSMIC's Large-Scale Search for Technosignatures during the VLA sky Survey: \\
Survey Description and First Results}

\author[0000-0002-4409-3515]{C.D. Tremblay}
\affiliation{SETI Institute, 339 Bernardo Ave, Suite 200, Mountain View, CA 94043, USA}
\affiliation{Berkeley SETI Research Center, University of California, Berkeley, CA 94720, USA}
\author[0009-0002-0688-513X]{J.Sofair}
\affiliation{Lafayette College, Department of Physics, Easton, PA 18045}
\affiliation{National Radio Astronomy Observatory, 1003 Lopezville Rd., Socorro, NM 87801, USA}
\author[0000-0002-9512-5492]{L. Steffes}
\affiliation{University of Wisconsin–Madison, Department of Astronomy, 475 N Charter St, Madison, WI 53703, USA}
\affiliation{Berkeley SETI Research Center, University of California, Berkeley, CA 94720, USA}
\author[0000-0003-0804-9362]{T. Myburgh}
\affiliation{Mydon Solutions (Pty) Ltd., 102 Silver Oaks, 23 Silverlea Road, Wynberg, Cape Town, South Africa, 7800}
\affiliation{SETI Institute, 339 Bernardo Ave, Suite 200, Mountain View, CA 94043, USA}
\author[0000-0002-8071-6011]{D. Czech}
\affiliation{Berkeley SETI Research Center, University of California, Berkeley, CA 94720, USA}
\author[0000-0002-6664-965X]{P.B. Demorest}
\affiliation{National Radio Astronomy Observatory, 1003 Lopezville Rd., Socorro, NM 87801, USA}
\author[0009-0001-8677-372X]{R.A. Donnachie}
\affiliation{Mydon Solutions (Pty) Ltd., 102 Silver Oaks, 23 Silverlea Road, Wynberg, Cape Town, South Africa, 7800}
\affiliation{SETI Institute, 339 Bernardo Ave, Suite 200, Mountain View, CA 94043, USA}
\author[0000-0002-3430-7671]{A. W. Pollak}
\affiliation{SETI Institute, 339 Bernardo Ave, Suite 200, Mountain View, CA 94043, USA}
\author[0000-0002-9473-9652]{M. Ruzindana}
\affiliation{Berkeley SETI Research Center, University of California, Berkeley, CA 94720, USA}
\author[0000-0003-2828-7720]{Siemion, A.P.V.}
\affiliation{SETI Institute, 339 Bernardo Ave, Suite 200, Mountain View, CA 94043, USA}
\affiliation{Breakthrough Listen, University of Oxford, Department of Physics, Denys Wilkinson Building, Keble Road, Oxford, OX1 3RH, UK}
\author[0000-0003-2669-0364]{S.S. Varghese }
\affiliation{SETI Institute, 339 Bernardo Ave, Suite 200, Mountain View, CA 94043, USA}
\author[0000-0001-7057-4999]{S. Z. Sheikh}
\affiliation{SETI Institute, 339 Bernardo Ave, Suite 200, Mountain View, CA 94043, USA}
\affiliation{Berkeley SETI Research Center, University of California, Berkeley, CA 94720, USA}

\begin{abstract}
Developing algorithms to search through data efficiently is a challenging part of searching for signs of technology beyond our solar system. We have built a digital signal processing system and computer cluster on the backend of the Karl G. Jansky Very Large Array (VLA) in New Mexico in order to search for signals throughout the Galaxy consistent with our understanding of artificial radio emissions. In our first paper, we described the system design and software pipelines. In this paper, we describe a postprocessing pipeline to identify persistent sources of interference, filter out false positives, and search for signals not immediately identifiable as anthropogenic radio frequency interference during the VLA Sky Survey. As of 01 September 2024, the Commensal Open-source Multi-mode Interferometric Cluster had observed more than 950,000 unique pointings. This paper presents the strategy we employ when commensally observing during the VLA Sky Survey and a postprocessing strategy for the data collected during the survey. To test this postprocessing pipeline, we searched toward 511 stars from the $Gaia$ catalog with coherent beams. This represents about 30 minutes of observation during VLASS, where we typically observe about 2000 sources per hour in the coherent beamforming mode. We did not detect any unidentifiable signals, setting isotropic power limits ranging from 10$^{11}$ to 10$^{16}$\,W.

\end{abstract}


\section{Introduction} \label{sec:intro}
The place of humanity in the Universe and the existence of life is one of the most profound and widespread questions in astronomy and society in general. Throughout history, humans have marveled at the starry night sky. It is intriguing to consider the distances to the nearest stars, the origins of our Universe, and how life on Earth came into existence. During the past 60 yr, the active search for life on other planets has not resulted in any evidence that it has occurred (e.g. \citealt{Ma_2023,Sheikh_2021,BLC1}). Even the long-debated ``Wow" signal \citep{wow} is no longer thought to be related to extraterrestrial life (\citealt{Wow_2024} and references therein). Therefore, we continue to wonder whether life is truly confined to this small planet orbiting an ordinary main-sequence star.

Starting in 2022, we built the Commensal Open-source Multi-mode Interferometric Cluster (COSMIC; \citealt{hickish_cosmic,tremblay_cosmic}), offering the first Ethernet-based digital framework on the world-leading Karl G. Jansky Very Large Array (VLA) in New Mexico for rapid, real-time analysis of astronomy data. COSMIC was designed to provide an autonomous real-time pipeline for observing and processing data for one of the largest experiments in the search for extraterrestrial intelligence to date. To meet this design goal, we have constructed a digital signal processing system and computer cluster that takes a copy of each of the digitized antenna voltages, corrects the signals, calibrates the gains, coherently beamforms on targets of interest, forms an incoherent sum, and then searches for signals of a spectral and temporal nature consistent with our understanding of artificial technological signals ($\sim$Hz wide emission signals; technosignatures). The output of this search is a database of ``hits" and small cutouts of the phase-corrected voltage data for each antenna around the hits called ``postage stamps." 

This is an opportune time for COSMIC to become operational, as it is possible to observe concurrently with the VLA Sky Survey (VLASS; \citealt{VLASS}). VLASS observes the entire northern hemisphere over two semesters in the VLA B-configuration, where Epoch 3 of the survey started in January 2023. Since March 2023, COSMIC has been operational and recording data from all 27 antennas with up to 1.2\,GHz of bandwidth. The survey continued recording data from the second half of the sky starting in May 2024 to complete a full census at declination above --40\,degrees. Observing along with this survey allows us to conduct one of the largest technosignature searches ever attempted within the frequency range of 2--4\,GHz, covering almost a million of stars within two years.

The vast amount of data generated by the software pipelines running on COSMIC and other similar backends has led to the development of fast and efficient methods for locating narrowband and Doppler drifting signals. In \cite{sheikh2019}, the authors discuss this problem in detail and conclude that the tree summation algorithm (also known as the Taylor tree method) is the most commonly implemented in modern technosignature searches due to its efficiency. However, the FastDD method \citep{Houston_2023} offers some promising improvements. The main software program used by the community employing the tree summation method to date is \textsc{TurboSETI} \citep{Enriquez_2019TT}\footnote{\url{https://github.com/UCBerkeleySETI/turbo_seti}; \url{https://ascl.net/1906.006}}, which is used on observations from the Robert C. Byrd Green Bank Telescope (e.g. \citealt{Gajjar_2021,Choza_2023}), Parkes 64m Telescope (e.g. \citealt{Price_2020,Sheikh_2021}), the Five-hundred-meter Aperture Spherical radio Telescope (e.g. \citealt{Wang_2023,Tao_2023}), and the Low-Frequency Array \citep{Johnson_2023}. 


Within the pipelines running on COSMIC, we use \textsc{seticore}\footnote{\url{https://github.com/lacker/seticore}}, which is a GPU-accelerated implementation of \textsc{TurboSETI} written in C++ with additional features: it includes postage-stamp extraction from the phase-corrected voltages of each antenna and visualization tools. This software is also implemented in the commensal technosignature search experiment on the MeerKAT telescope (Czech et al. in prep.). As with \textsc{TurboSETI}, the user input parameters include the signal-to-noise ratio (SNR) threshold and the Doppler acceleration drift rate\footnote{For a description of how the noise statistics are calculated and the software's limitations see \cite{Choza_2023}}. It is possible to determine the precise expected drift rate based on the planet's orbital motion \citep{Li_2022,LiM_2023}, but this is only applicable to targeted surveys of known planetary systems. When detecting signals from nearby stars where planets are likely to exist but have not yet been discovered, as with COSMIC, broad search criteria are used. To search the data, we use an SNR threshold of 8 and a Doppler acceleration range of $\pm$ 50\,Hz\,s$^{-1}$. This limit is much broader than necessary for known exoplanets \citep{sheikh2019,Li_2023} but accounts for a broader range of possible scenarios where technology is orbiting off the planet but within the solar system (like Earth's satellites). Coherently and incoherently beamformed data are searched using these parameters, with a frequency resolution of 7.63\,Hz per channel for the total 8-second recording segment, resulting in a drift rate bin size of 0.95\,Hz\,s$^{-1}$. We also collect and save data with a drift rate of exactly 0\,Hz\,s$^{-1}$. Such data are often discarded in many experiments as a way to limit the intrusion of terrestrial signals into the data. However, the 0.95\,Hz\,s$^{-1}$ may not provide the necessary resolution to detect signals arriving from systems similar to our nearest neighbors, such as TRAPPIST-1g  (1.33\,Hz\,s$^{-1}$) or Kepler-438 b (0.7090\,Hz\,s$^{-1}$), as calculated by \cite{Li_2022}. 

Once we have these detected signals and their associated metadata collected into a database, we need to devise plans to sort through these data to eliminate false positives likely associated with terrestrial radio emitters or instrumental artifacts and find new signals with astronomical origins. Although many of the experiments published so far involved manually investigating the dynamic spectra of the most promising events, this is no longer practical where the COSMIC pipeline identifies millions of signals per hour over thousands of sources. Although observational changes such as looking at signals that traverse across a multibeam receiver (i.e. \citealt{Huang_2023}) or flagging known terrestrial signals within the real-time pipeline could be considered in the future, other ways of evaluating the data are required to investigate the data collected in the first six months of operation with COSMIC on the VLA.

In this paper, we describe a postprocessing strategy for the data collected and processed by COSMIC during VLASS observations. In Section 2, COSMIC's observations during VLASS are summarized, and the observations that have been evaluated are described.  
In Section 3, we discuss the effects of a radio frequency interference (RFI) excision technique used after the data have been collected on COSMIC. In Sections 4 \& 5 we discuss a postprocess for the evaluation of COSMIC-generated data for SETI and the first results from one of the VLASS observation nights. Finally, we discuss the impact of the results and conclusions of this work.

\section{Observations}
The observations described in this paper are based on data taken simultaneously with other VLA observing projects but analyzed separately, following the relevant National Radio Astronomy Observatory (NRAO) policies\footnote{\url{https://science.nrao.edu/observing/proposal-types/commensal-observing-with-nrao-telescopes}} for commensal observation. Therefore, the COSMIC team does not control the observational strategy or calibration cadence. Our processing and analysis do not impact the standard NRAO operation and analysis used by the primary observer or the NRAO staff.

\subsection{VLASS Observing Strategy}

During VLASS, we observed and recorded data while the VLA was in the B-configuration and the hybrid configuration called B-North-A. Two observation campaigns, one in 2023 and one in 2024, represent the third epoch of this survey, which covers 80 percent of the sky. As part of this survey, large sections of the sky are rapidly covered by the telescope using the on-the-fly (OTF) mosaic method\footnote{\url{https://science.nrao.edu/facilities/vla/docs/manuals/obsguide/modes/mosaicking}}. During this time, observations are broken into segments called “tiles”. Each tile covers between 25--40\,deg$^2$ by moving in right ascension (RA) at a rate of 3.3\,arc\,minutes per second, meaning there are approximately 5--10 seconds on each source before the source is outside the primary beam (the slew rate is slower at declination's below $\sim$--20 degrees). After scanning in RA for approximately 10 minutes, the telescope moves in declination by 7.2\,arc\,minutes and then scans again in RA covering the same distance as the previous track. Overall, it takes about 2 hours to observe each tile. For further details on the observation, see \cite{lacy_vlass}.

For COSMIC, we adopted an observing strategy where voltages are recorded in 8-second increments and processed separately. As described in \cite{tremblay_cosmic}, the targets on which to form coherent beams, phase-centered pointing toward particular sources with a field diameter equal to the point-spread function, were selected based on the sources likely to spend the most amount of time within the half-power point of the telescope's primary beam when a fully recorded field of view (FOV) is considered. The list of targets is from the technosignature target catalog, which itself is a selection of sources based on the $Gaia$ Space Telescope DR2 catalog \citep{Czech_2021}. For each two-hour observation of a single tile, the COSMIC software pipelines forms approximately 4000 coherent beams on sources in the target catalog.

\begin{figure}
\includegraphics[width=0.48\textwidth]{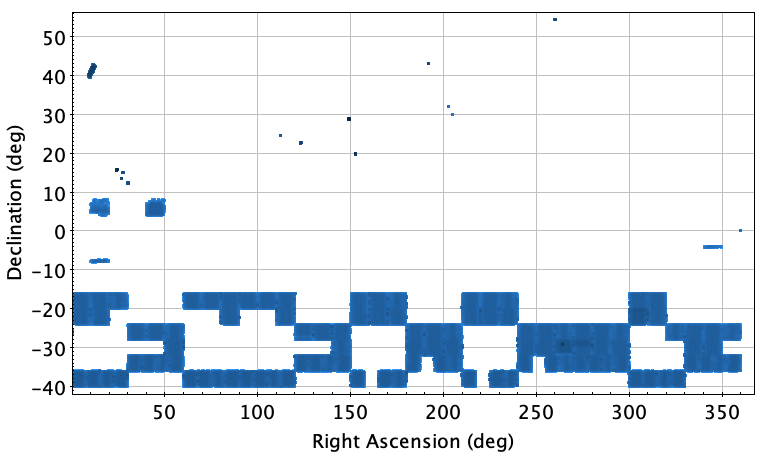}
\caption{A plot of all the coordinates of targeted stars during VLASS recordings from 25 March 2023 to 15 June 2023. All recorded data of scientific quality that are contained with the COSMIC database for Epoch 3.1 are represented here.  \label{fig:Beams}}
\end{figure}

During VLASS Epoch 3.1 in 2023, COSMIC observed and recorded science-ready data from 25 March 2023 to 15 June 2023, as shown in Figure \ref{fig:Beams}. We accumulated 326,504 unique target fields (unique RA and Dec coordinates), representing 267,510 stars towards which we formed coherent beams, and 359,353,639 ``hits" found by \textsc{seticore}.  To assess the strategy and data quality we used 30 minutes of data observed on 15 April 2023 during the recording of a partial VLASS tile. The VLA was in B-configuration and recording data at 2--4\,GHz in the 8-bit recording mode, and 24 of the 27 antennas were online and recording for the duration of the observation. Every 15 minutes, the VLA conducted phase calibration scans, with at least one scan of 3C 286 per tile used for flux calibration. 

\subsection{COSMIC Recording and Processing Setup}
\begin{figure*}
\includegraphics[width=0.98\textwidth]{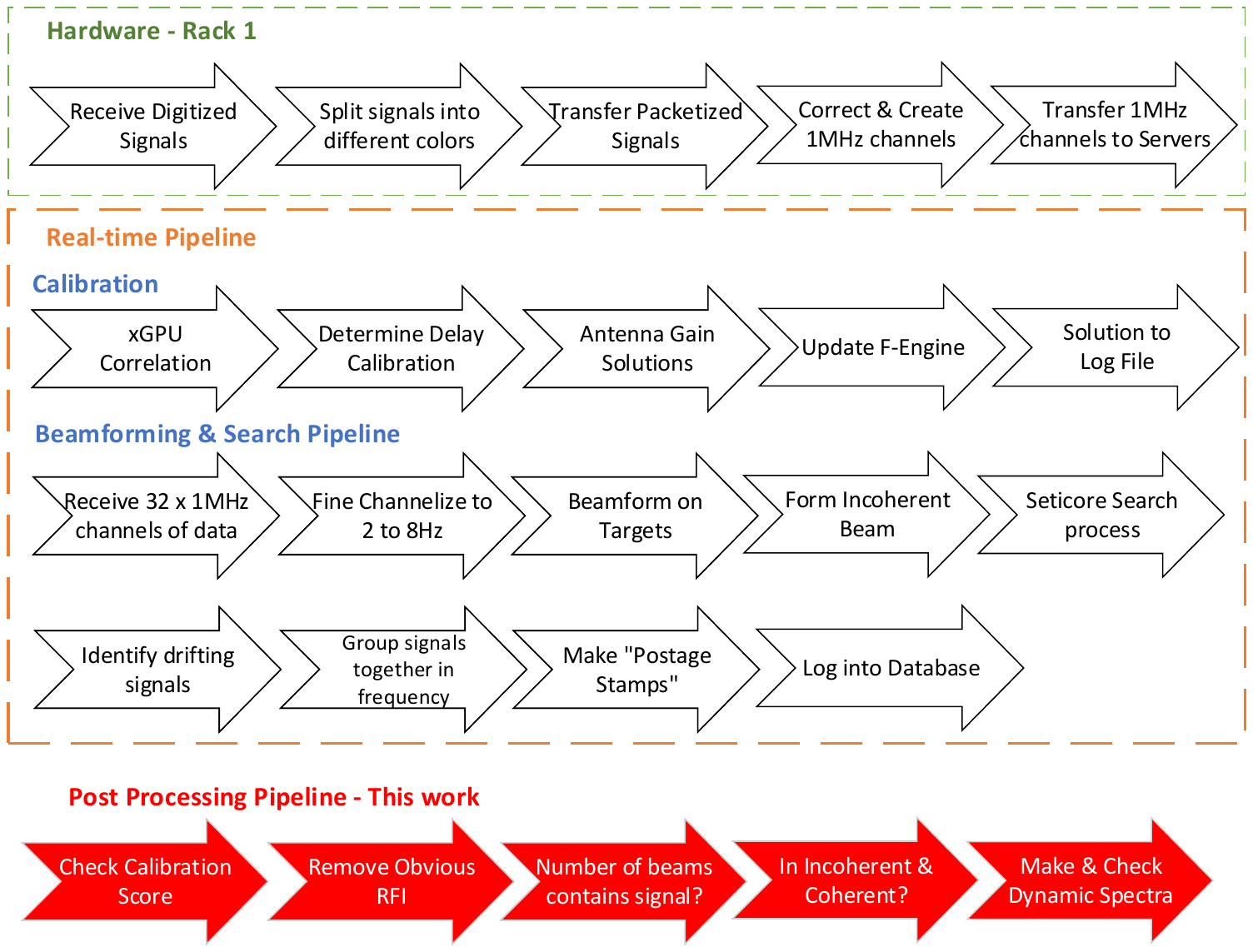}
\caption{Summary of the pipeline implemented in COSMIC. Based on the telescope's observational intent, the data will either flow from the hardware to the calibration pipeline or from the hardware to the search pipeline on the GPUs. The postprocessing step, which will be described later in this work, is a separate data flow and does not happen in real time. A more detailed depiction of the process can be found in Figures 3 and 8 of \cite{tremblay_cosmic}}. \label{fig:Pipeline}
\end{figure*}

COSMIC is divided into four distinct segments (digital signal processing, calibration, beamforming, and search), the last three of which take place in real time and are outlined in Figure~\ref{fig:Pipeline} and described in detail in \cite{tremblay_cosmic}. Digital signals are received and processed by the first rack of COSMIC. Each field programmable gate array (FPGA) board receives the digital signals from two of the antenna's digital transmission streams using existing VLA firmware provided by NRAO (but processes each independently). They also
\begin{itemize}
    \item compensate for delays due to both signal propagation from the astronomical source of interest to the VLA antennas and fiber optic signal propagation from the antennas to the COSMIC system;
    \item removes local oscillator offsets and creates coarsely channelized 1\,MHz channels from the broadband input;
    \item form packets of digitized coarse channels to transmit via a 100\,Gb Ethernet switch to a runtime-determined downstream processing node (the GPU node).
\end{itemize}

\begin{figure}
\includegraphics[width=0.48\textwidth]{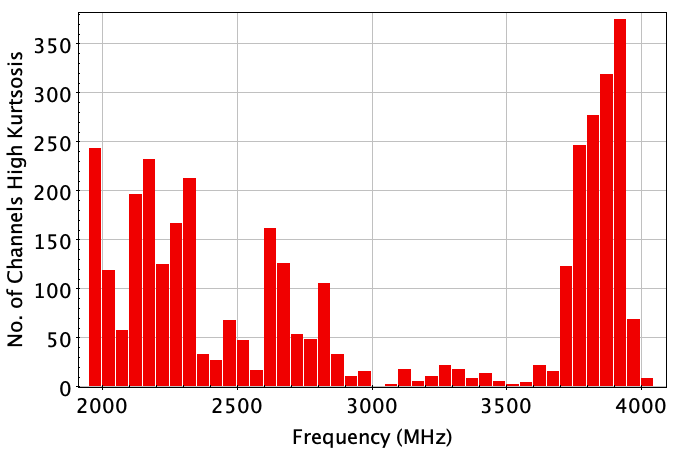}
\caption{Histogram showing the number of frequencies per frequency window that have high kurtosis, implying high levels of RFI. The edges at 2--2.5\,GHz and 3--3.5\,GHz show significant RFI. Therefore, for VLASS, the central frequency range of 2.5--3.5\,GHz was processed and searched.}
\label{fig:Full_Band}
\end{figure}

Incoming data from the VLA consist of two intermediate frequencies (IFs) of 1024\,MHz each, for a total of 2048\,MHz, when in the 8 bit recording mode. During the 2023 VLASS observations, we had up to 15 operational GPU servers (some shut down as we worked on system stability), each containing two processing nodes. In order to search for signals within the 960\,MHz of bandwidth, we used the 100\,Gb switch to transfer 32 $\times$ 1\,MHz coarse channels to each node, which were then split between the two IFs (480\,MHz each) between the 2 and 4\,GHz that the COSMIC system ingested. As shown in Figure \ref{fig:Full_Band}, there was significant RFI on the edges of the band, so we focused the processing and search on 2.5--3.5\,GHz, as will be further discussed in Section 3.

Each 32\,MHz of bandwidth is processed through either the calibration pipeline or the technosignature search pipeline. The VLA observations are set with a series of ``intents,"\footnote{\url{https://www.aoc.nrao.edu/~cwalker/sched/INTENTs.html}} which are assigned by the person creating the scheduling block. When the intent contains the word ``calibration" the data are processed through the calibration pipeline, as defined in the second line of Figure \ref{fig:Pipeline}. When the intent is ``OBSERVE TARGET," the data are run through the technosignature search pipeline. The following sections describe these processes in detail. The information about what the pipeline detected is recorded to an SQL database, and ``postage stamps" of the calibrated voltages are saved on disk. 

The data from the 15 April 2023 observations include coherent beamformed results on 511 unique $Gaia$ DR2 sources from the catalog created by \cite{Czech_2021}, some of which were searched for drifting signals multiple times when the declination strips overlapped with a previous slew of the telescope. We also searched within the incoherent sum in each 8\,seconds of data. 

\subsection{Calibration}
The COSMIC processing pipeline follows standard calibration procedures for radio astronomy data with a strong focus on delay and phase calibration. Currently, the automated pipeline does not conduct absolute flux density calibration. However, the data are saved on disk, and this can be applied to the data prior to reporting any values. The pipeline uses the \textsc{xGPU}\footnote{\url{https://github.com/GPU-correlators/xGPU/blob/master/README}} \citep{Clark:2013} software correlator to cross-correlate the raw voltages and create the visibility data products in four polarizations (RR, LL, RL, and LR, where R and L refer to right and left circular polarization, respectively), which are saved in $uvh5$ files\footnote{\url{https://pyuvdata.readthedocs.io/en/v1.5/_modules/pyuvdata/uvh5.html}} on the storage nodes. Once each of the 32\,MHz of data is correlated and the delays and gain solutions are computed, the data are collated back together across all frequencies on the COSMIC head (computer) node to be reorganized and reported. The new residual delay and phase solutions are uploaded to the FPGAs, and the resultant solutions are saved in a $json$ file format in a folder with the $uvh5$ file for posterity. 

As part of the calibration process, after the solutions are brought together and sorted by frequency, the solution quality is graded based on the stability of the phases. With each new calibration processing pipeline application, the FPGA signal corrections get updated. Therefore, each new calibration has had some calibration applied, and the results of the phase are really a difference from the previous calibration. The main grading criterion we use to define a good calibration compared to a bad calibration is phase stability. When the phases are flat across frequencies (per each polarization and antenna), the resultant grade is given the value of 1. When the phases are chaotic, uncorrelated, or contain significant wraps, the grade value decreases toward 0. See Equation 9 in \cite{tremblay_cosmic} for details on how the overall grade is computed in detail.

Uncorrelated individual frequency channels or antennas, often caused by an incorrect delay model or the antenna still slewing during recording, can corrupt the overall grade value. That is why, in addition to the overall grade, we compute a grade per frequency channel across antennas (Equation (8) in \cite{tremblay_cosmic}) and a grade per antenna across all frequency channels (Equation (7) in \cite{tremblay_cosmic}) to better isolate these instances. However, these grades are determined before the calibration is applied and representative of a differentiation between calibration runs more than an accurate measure of the determined calibration values for the current observation.\footnote{Current automated processes on COSMIC generate a predictive overall grade from the calibration gain solution that is used to predict phase stability in subsequent target observations. However, this was not implemented at the time the data evaluated in this paper were written.}

\begin{figure}
    \centering
    \subfigure[]{\includegraphics[width=0.48\textwidth]{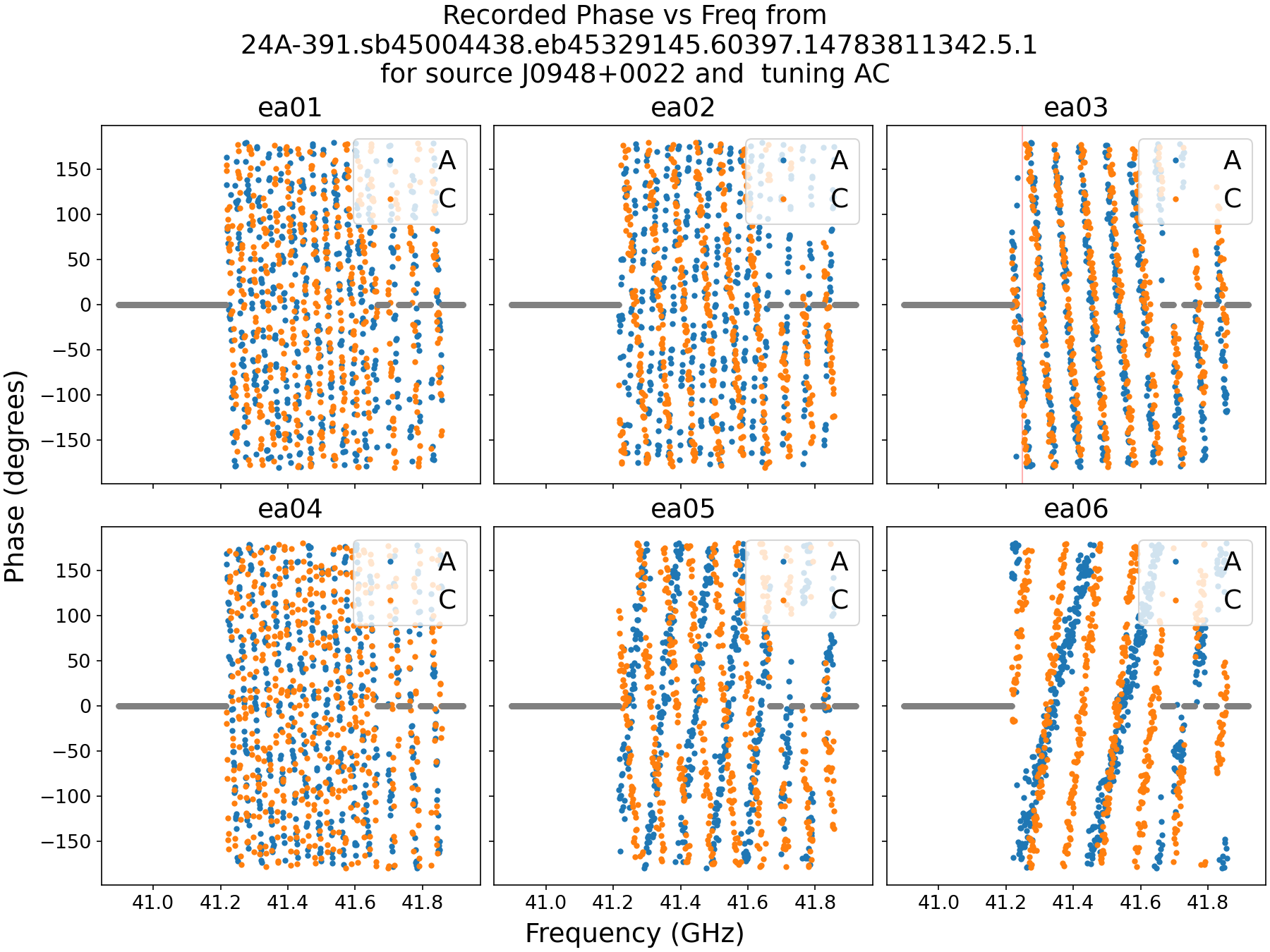}}
    \subfigure[]{\includegraphics[width=0.48\textwidth]{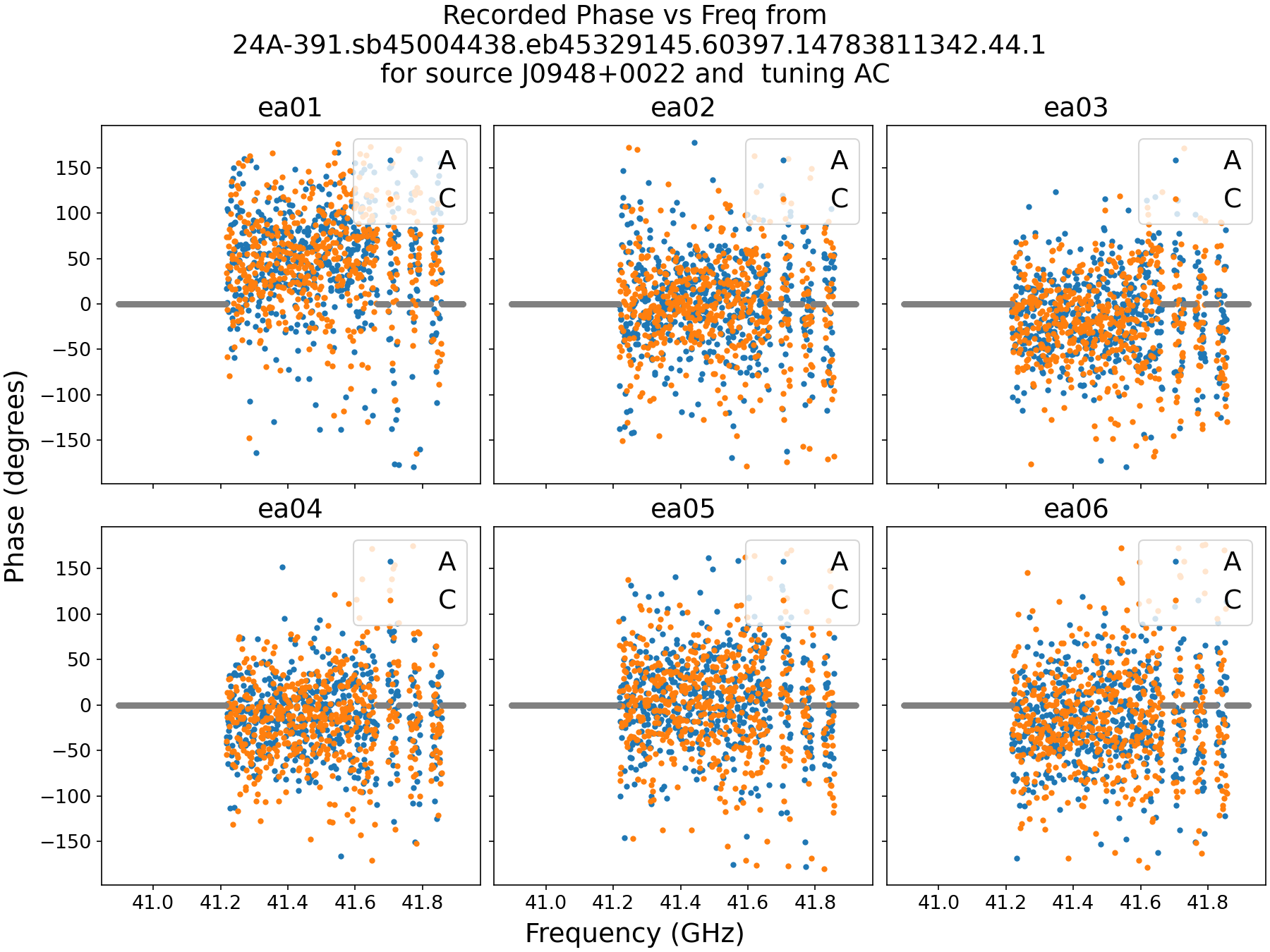}}
    \subfigure[]{\includegraphics[width=0.48\textwidth]{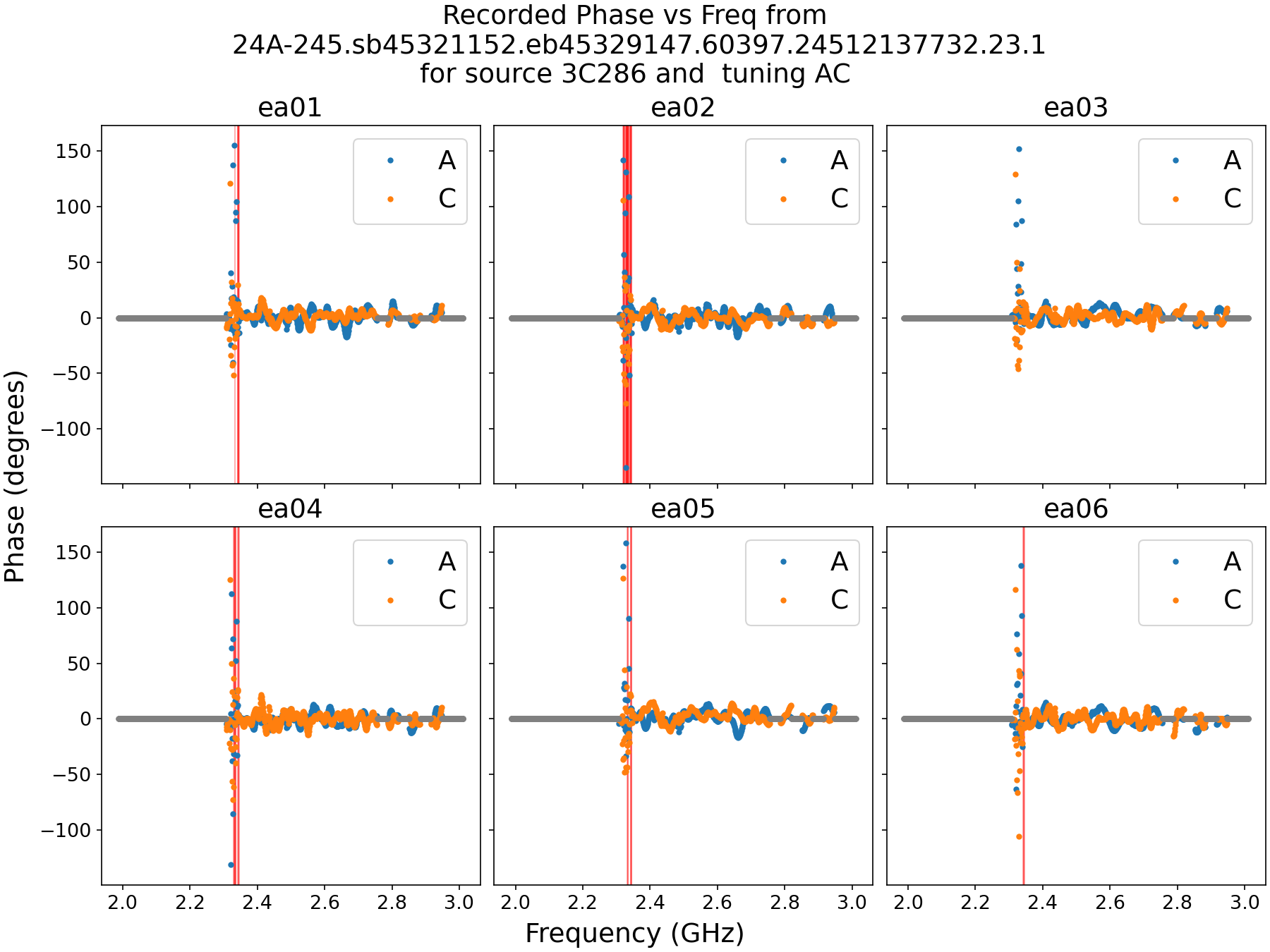}}
\caption{Example calibration plots from the AC tuning generated by plotting the phases prior to the calibration being applied. In (a), the overall grade is $<$0.2, (b) has a score of 0.639, and (c) has a score of 0.991. A score greater than 0.6 is considered a ``good" calibration, as this represents a score prior to the correction to the FPGA. Therefore, it shows the ease with which a solution can be found more than the actual solution. The red vertical strips indicate 1\,MHz channels that are flagged as containing RFI signals during the calibration run. \label{fig:Cal_plots}}
\end{figure}

Examples of the calibration phases plotted as phase degrees versus frequency are shown in Figure \ref{fig:Cal_plots}. The overall calibration grade and the plots, along with other diagnostic plots, are sent to a \textsc{Slack}\footnote{\url{https://slack.com/intl/en-gb/}} channel for posterity and to allow those reviewing the data for scientific use a rapid review and evaluation of the calibration quality during the observations of interest. When a calibration grade is above a value of 0.6, representing that there is no more than a 40\% drop in sensitivity, and chosen as value suggestive that all but one or two antennas have phases that are easy to calibrate, no further investigation of the solutions is completed. The resultant science data proceeding the calibration are taken as worthy of investigation. When the score is less than 0.6, the plots and other diagnostics are reviewed to see if the data are scientifically valid. These phase solutions are critical; in coherent beamforming, improper phase solutions will cause the coherent beam to be offset from the position of the target source, creating either a significant reduction in sensitivity or a smeared point-spread function (depending on the severity of the problem).

In the last year, we improved the phase solution calculation by flagging for 1\,MHz channels subject to narrowband RFI. For the spectrum of a given tuning, baseline, and polarization averaged over time, a channel that satisfies

\begin{equation}
    |f_i - \median{f}| \geq 5\times|\mad{f}|
\label{eq:rfiflagging}
\end{equation}

is considered to contain RFI, where $f$ is the frequency, $i$ the channel index, and $MAD$ the \textit{median absolute deviation}.
A flagged spectral value $f_i$ is replaced with the $\median{f}$ so as not corrupt the calculated gains and, in turn, the applied phase solution. We note that while I RFI emission is largely coherent, it is not usually continuous, and so deriving gains from RFI-contaminated signals will likely lead to gain solutions for that channel that do not accurately represent the scientific data in the target observation.

For longevity and retrospective analysis, two calibration tables in the COSMIC database are created. They are `AntennaCalibration' and `ObservationCalibration.'
In the `ObservationCalibration' table, we store the overall calibration details linked to the calibration observation. This is mostly used for grade performance evaluation in time studies. `ObservationCalibration' stores the calibration observation ID, the reference antenna, the overall grade, and the output directory for the calibration observation and its products.
`AntennaCalibration' stores the antenna and tuning-specific calibration details, which are used to further inspect irregularities noted in grade evaluation studies. For each antenna entry, while linked to the calibration observation ID, we store the antenna name, tuning, number of coarse channels processed in the calibration run, and number of RFI-flagged channels during the calibration run.

\begin{figure}
\includegraphics[width=0.5\textwidth]{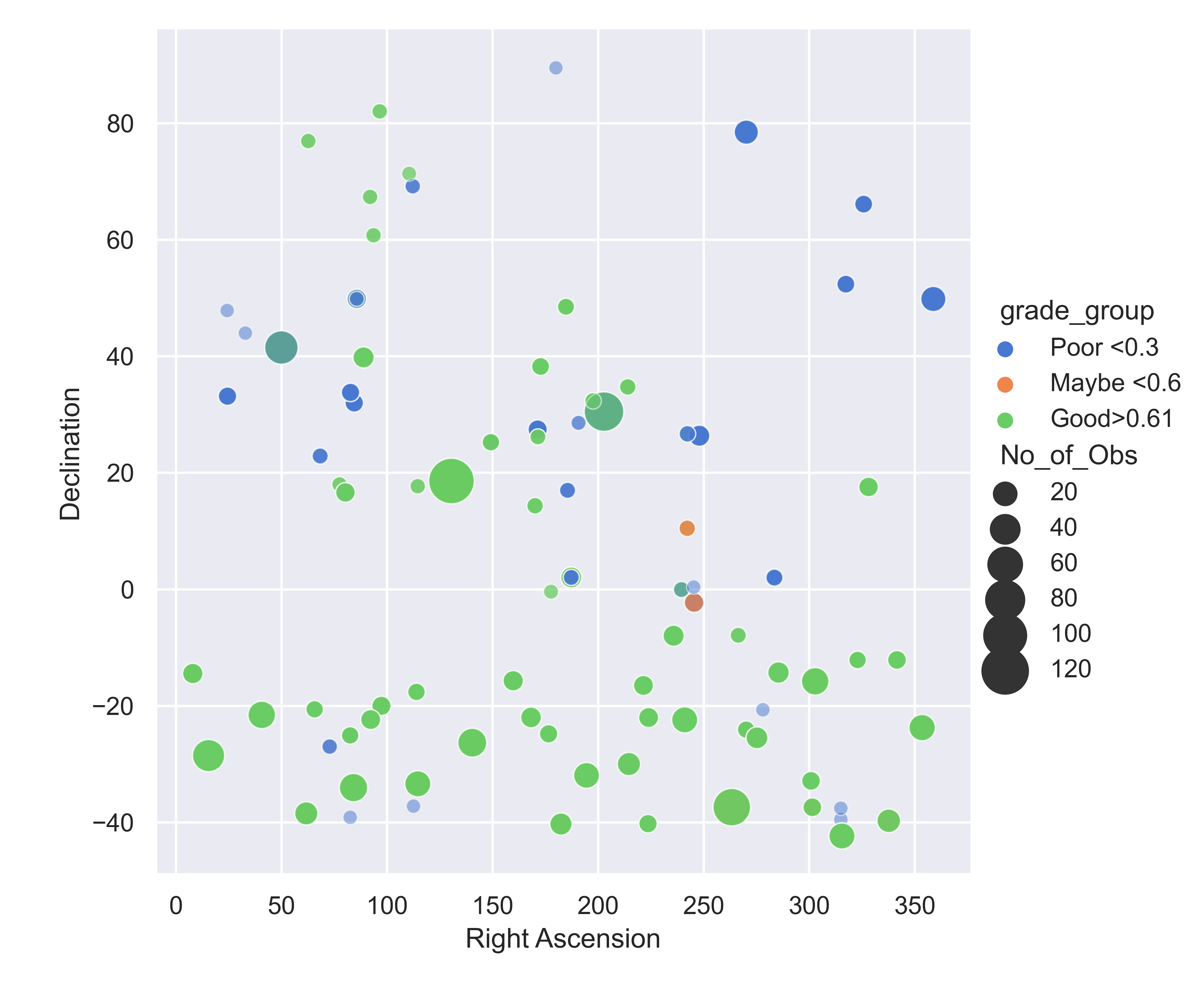}
\caption{Calibration results for 2 months of operation.  The colors of the bubbles represent the grade of the calibration, where the higher value is a lower SNR across the phase and the size of the bubble represents the number of times the source was observed. As can be seen, the calibration is good most of the time. \label{fig:Cal_score}}
\end{figure}

To evaluate the stability of the calibration score and COSMIC's pipeline operation, we plotted the data from May and June 2023. The most common calibrators used by observers and recorded by COSMIC were J2355+4950, J1024-0052, J1033+6051, and 3C286. As demonstrated in the plot in Figure \ref{fig:Cal_score}, more than 80\% of the calibrations across a range of frequencies have a score of greater than 0.6. In particular, the best scores are often from strong phase and flux calibrators used frequently in the calibration of VLA data. Many of the poor-quality data plots (in blue) are often from weak phase calibrators chosen by the observer or due to a frequency change with respect to previous observations. 

For the observations reviewed here from 15 April 2023, the grade of the observations varied significantly but most were in the range of 0.4--0.9. All corrected gain plots show good phase stability for at least 24 antennas. 

\subsection{Beamforming during OTF Scans}
Although the beamforming code and science verification are discussed in depth in \cite{tremblay_cosmic}, here we discuss their application when the telescope is using the OTF scanning mode as used during VLASS. The COSMIC pipeline uses the beamforming code ``Breakthrough Listen Accelerated DSP Engine" (BLADE\footnote{\url{https://github.com/luigifcruz/blade}}), which was developed as part of the refurbishment of the Allen Telescope Array in California, USA.

The data recorded during an OTF scan are partitioned into files spanning 8.338608s each. The OTF observation system ensures a constant phase across each file by instructing the antenna to only update phase centers on file boundaries. This limit is also determined as the upper limit after which beamforming and dedoppler search cannot be achieved within ``real time." The coordinates of the beams formed are sourced from candidates within the half-power point of the FOV around the phase center as received from the target selector system (see Figure 16 in \citealt{tremblay_cosmic}). Consequently, the beam coordinates are independently derived from each 8s partition of the data. Before the actual beamforming computation, each coarse channel is upchannelized to the requested resolution without reducing the number of fine spectra below 50 (typically 64 time steps for VLASS observations with an FFT size of 131,072), to the benefit of the dedoppler search, and while keeping the overall processing time within ``real-time" bounds. The upchannelization exhausts all time samples for the coarse channel before iterating to the subsequent coarse channel. The following beamforming computation processes each fine spectrum. We collect the whole time span before we do a dedoppler search. Therefore, the dedoppler search is computed iteratively across each upchannelized, beamformed coarse channel, repeating for each time partition of the OTF scan. 

\begin{figure*}
\centering
\includegraphics[width=0.95\textwidth]{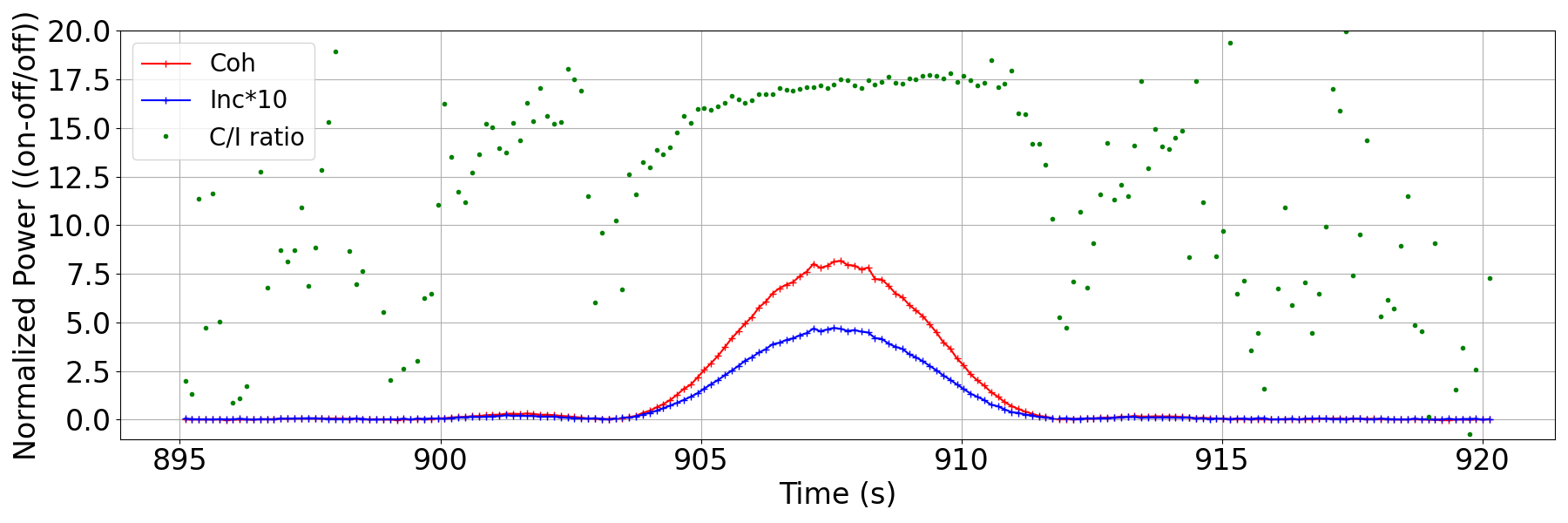}
\caption{The plot above shows the power over time for the frequency channel containing the peak intensity of the W51 6.7\,GHz methanol maser emission. The shape of the line is expected to correspond to the antenna's primary beam response. As a comparison, the power response of coherent and incoherent beams (multiplied by a factor of 10 to improve visibility) is plotted, along with their ratio. The ratio is ideally expected to equal the number of antennas used in the summation, which was 21 in this test. Imperfect calibration reduces the ratio away from the ideal value. The high scatter of the ratios away from the peak is due to large errors from small, noisy values.}
\label{timeseries}
\end{figure*}

To ensure the beamformer coefficients are updated appropriately, an observation of the methanol maser emission at the star-forming region W51 (e.g. \citealt{Etoka_2012}) was taken on 31 May 2023 using the same OTF parameters as VLASS and 3C286 was used to calibrate the data through the standard pipeline. By plotting the time samples as a function of normalized beam power for the frequency channel that contains the peak emission of the maser, we see a pattern that mimics the primary beam response of the VLA dishes as expected, shown in Figure \ref{timeseries}\footnote{This observation was taken during test time with 21 antennas online.} \citep{PB_VLA}. 

\subsection{Real-time Search Pipeline}
Each 8s segment of recorded data is searched for narrowband signals following beamforming. For this, we use the software package {\sc seticore}, which uses the same Taylor tree method as {\sc Turboseti} but is rewritten for GPU acceleration. In the pipeline running COSMIC during VLASS, {\sc seticore} searches for narrowband signals that have an SNR above ten and across a drift rate of $\pm$50\,Hz\,s$^{-1}$. For each signal it detects, the metadata about the signal characteristics are logged into an SQL database, and a segment of the calibrated and channelized voltages is saved as ``postage stamps."

\begin{figure}
    \centering
    \includegraphics[width=0.95\linewidth]{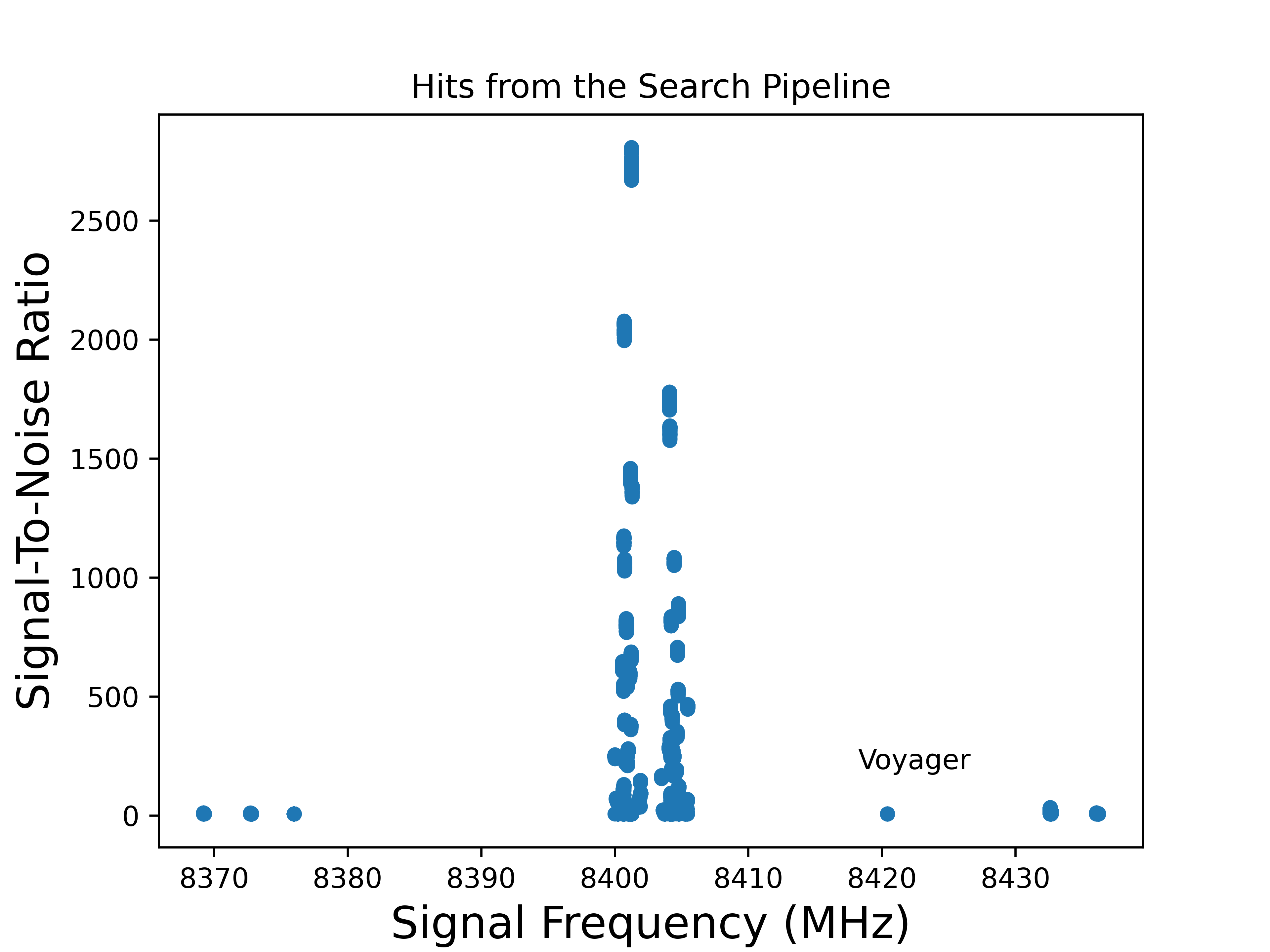}
    \caption{A plot of the signal-to-noise ratio versus observing frequency showing all of the hits found in the observations toward Voyager detected by {\sc seticore}.}
    \label{voyager}
\end{figure}

To test the efficiency of the real-time pipeline at detecting technosignatures, we pointed the telescope toward the Voyager I space satellite using VLASS-style observations. We then let the real-time pipeline run with no input except to prioritize a coherent beam toward Voyager's location. As shown in Figure \ref{voyager}, {\sc seticore} recorded information on 4121 signals. Voyager's signal around 8420\,MHz was one of these signals detected. We note that additional testing will be discussed in a future publication with the full VLASS data release.

\section{Radio Frequency Interference}
As mentioned in Section \ref{sec:intro}, the COSMIC postprocessing pipeline searches for signals that are similar to anthropogenic sources of radio emissions in their spectral and temporal signatures. Despite providing a valuable reference for the characteristics of technosignatures, human-made radio emissions are also one of radio astronomy's greatest hindrances. Given the ubiquity of anthropogenic radio emission technology (e.g., cellular networks, satellite radio, etc.), any radio telescope located near enough to humans will detect these unwanted radio signals. For COSMIC, this means signals that do not originate from the target source will be picked up and flagged as potential technosignatures. The signals that cause these false positives are known as RFI. We therefore have created a software package described below to identify prominent and persistent sources of RFI during calibrator observations. The assumption here is that the signals detected in a short observation toward the calibrator are all from sources of terrestrial radio emission and not signals of interest. 

\subsection{Classification and Identification}
Two characteristics of the technosignatures we are seeking are that they could be narrowband (signal contained in $\sim$Hz wide channels) and that their frequency would change over time as a result of the Doppler effect \citep{2019ApJ...884...14S, Li_2022}. Technosignatures can be distinguished from typical astronomical sources by their widths in frequency space. The finest-frequency astronomical emitters discovered thus far are found to produce signal widths on the order of $10^{2}$ Hz \citep{gray99, zhang_fast}.


RFI permeates astronomical data, causing narrowband spikes in detected power with time-varying intensities and frequency instabilities\footnote{\url{https://science.nrao.edu/facilities/vla/docs/manuals/obsguide/rfi}}$^{,}$\footnote{\url{https://www.parkes.atnf.csiro.au/observing/rfi/how_does_rfi/how_does_rfi.html}}. The similarities between terrestrial radio signals and technosignatures make it challenging to determine if the source of a signal is of extraterrestrial origin or not. Contemporary technosignature search algorithms therefore output a large number of false positives. Thus, the search for an extraterrestrial needle in a ``cosmic haystack" \citep{Wolfe_1981,wright_cosmic_haystack} becomes exponentially more challenging in the presence of RFI. Even if a signal can be confirmed to be anthropogenic, its highly variable properties make it troublesome to remove.


Currently, technosignature searches employ a variety of RFI removal techniques, including observational strategies and postobservation data analysis. Observational techniques include the use of cadenced-based observation (some time pointing away from the primary source) for single-dish telescopes (e.g. \citealt{Price_2020}), and the analysis of which beams---coherent and/or incoherent---a signal is found in for interferometers (demonstrated in this work). However, multitelescope follow-up is also a viable and suggested method for signal confirmation (i.e. \citealt{Ma_2023,Johnson_2023}). The identification of terrestrial radio signals includes the analysis of $\chi^{2}$ statistics \citep{weber_97}, the use of machine learning \citep{zhang_fast, ma_ts_deeplearning}, RFI subtraction using a reference signal \citep{briggs_00}, and much more. Many of the techniques that fit into the latter category are mathematically intense and/or computationally expensive. Here we seek to develop a new method to swiftly flag frequency ranges that are heavily tainted by RFI to allow for more efficient data analysis.

\subsection{RFI Identification with Excess Kurtosis}
Our first line of defense against RFI uses the statistical measure of excess kurtosis, $exkurt$, which gives the measure of the infrequent outliers in a distribution. We use the Fisher\footnote{\url{https://docs.scipy.org/doc/scipy/reference/generated/scipy.stats.kurtosis.html}} definition of kurtosis, $kurt$, in this work, given by
\begin{equation}
  exkurt = \frac{\mu_4}{\mu_2^2} - 3,
\end{equation}

where $\mu_4$ is the 4$^{th}$ central moment of the distribution and $\mu_2$ is the 2$^{nd}$ central moment (variance). Under this definition, a standard normal distribution has an excess kurtosis of 0.

The astronomical noise in our data is expected to have power normally distributed in any given frequency bin due to the incoherence of the electromagnetic waves collected by the VLA \citep{smith_sk}. Therefore, large enough deviations from Gaussianity imply the presence of a signal. We can use this to our advantage to rapidly flag frequency bins that are dominated by RFI, or ``dirty." To filter out the false positives in the database during data postprocessing, we developed a new \textsc{Python} package called the Categorization of RFI in COSMIC with Kurtosis for Extraterrestrial Searches (\textsc{CRICKETS}\footnote{\url{https://github.com/jareds-laf/CRICKETS}}). 

The use of excess kurtosis to flag RFI-dominated frequencies is not a novel technique. \cite{nita_sk} proposed the spectral kurtosis estimator, $\widehat{SK}$, a mathematical object used to flag RFI within the context of radio astronomy. This estimator was generalized by \cite{nita_sk_generalized} and implemented in \textsc{hyperSETI}\footnote{\url{https://github.com/UCBerkeleySETI/hyperseti}}, a \textsc{Python} package designed to search spectral data for technosignatures in both single-dish and interferometric data. COSMIC currently runs a different technosignature search package (\textsc{seticore}) that does not have a robust RFI flagging method. The lack of this key function in the COSMIC beamforming and search real-time pipeline is the primary motivation behind the use of \textsc{CRICKETS} within the pipelines running on COSMIC.

\subsection{\textsc{CRICKETS} Application}
Frequency, power, and time data are fed to \textsc{CRICKETS} in the form of a $filterbank$ file generated from forming an incoherent beam on a VLA observation toward a calibrator source. This was done with the assumption that no technosignatures would be in the field surrounding the calibrator and that the emission detected in the data would be dominated by broadband emission from the bright calibrator source and RFI. \textsc{CRICKETS} reads these $filterbank$ files in using the \textsc{Blimpy}\footnote{\url{https://github.com/UCBerkeleySETI/blimpy}} \textsc{Python} package. Once the file is loaded in, the power is averaged over the time axis and tabulated with their respective frequencies.

The next operations that \textsc{CRICKETS} performs are to organize the data to a user-specified number of frequency bins, rescale the data\footnote{The data is re-scaled to avoid any infinities in the excess kurtosis calculations.} by a factor of $10^{9}$, and to calculate the excess kurtosis (Equation 1) of every frequency bin using \textsc{Scipy}. The excess kurtosis calculation is computed using all of the fine-frequency channels within a given bin. Bins that are dominated by RFI are flagged based on their excess kurtosis relative to the excess kurtosis threshold, which is the second user-specifiable quantity in the analysis. Bins can have an excess kurtosis of a certain magnitude before being flagged as dirty if they exceed the threshold. Bins flagged as dirty are compiled and output to a $CSV$ file using \textsc{Pandas}. 

\subsection{S-band RFI Determination and Results}\label{sec:crickets_results}
\begin{figure}
    \centering
    \includegraphics[width=0.98\linewidth]{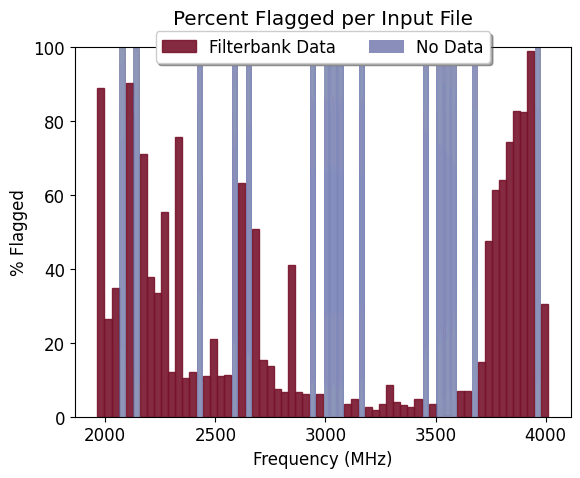}
    \caption{Percentage of frequencies flagged per filterbank file. Each dark red rectangle represents a filterbank file. Each dark blue rectangle represents a frequency range within the S-band for which we did not have data.}
    \label{fig:filfil_results}
\end{figure}

We used \textsc{CRICKETS} to analyze 48 $\times$ 32\,MHz subbands covering the S-band (2--4\,GHz) non-continuously. We generated an incoherent sum using \textsc{BLADE}\footnote{\url{https://github.com/luigifcruz/blade}} version 0.7.4 toward the source 3C286 (RA: 4h9m22s, Dec: 12d17m39.8475s) on May 10th, 2023 with the VLA in the B-configuration. We analyzed the data observed during Epoch 3.1 of VLASS using the tool \textsc{CRICKETS} to determine the number of frequency bins and excess kurtosis threshold that could be applied to the entire sample of data. \footnote{In the future, regular scans will be done to look for the persistent nature of RFI and find sources that change as a function of time and elevation.} Generating the time-averaged power spectra and excess kurtosis plots revealed that 256 bins per file (corresponding to 125\,kHz per bin or 15,625 fine channels per bin) with an excess kurtosis threshold of 5 was optimal. In some files, regions dominated by RFI went unflagged, but if the threshold was lowered even slightly, regions in other files without RFI would have been falsely flagged. With the parameters we chose, about 29.88\% of the frequencies evaluated were flagged as dirty (see Figure \ref{fig:filfil_results}).

\begin{figure}
    \centering
    \includegraphics[width=0.98\linewidth]{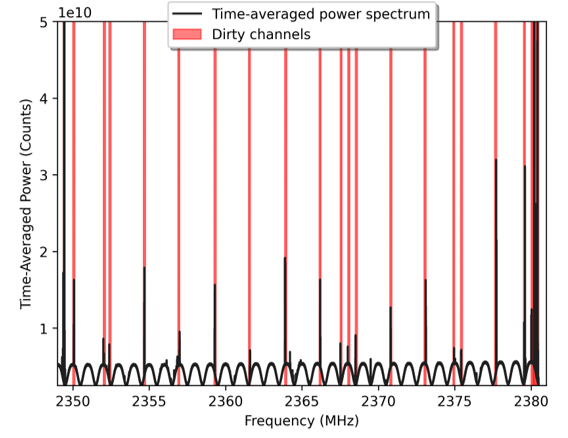}
    \caption{Example of a time-averaged power spectrum in the ~2348.5--2380.5 MHz band. The transparent red regions mark the locations where CRICKETS detected RFI. The fine-frequency spikes in time-averaged power are a result of RFI. We know this to be true because these data come from observations of the calibration source 3C286. The sinc-like pattern at the bottom of the plot is a result of the polyphase filterbank (PFB) process, which is a necessary step in COSMIC's signal processing procedures. This will be discussed further in Section \ref{sec:crickets_results}.}
    \label{fig:crickets_example}
\end{figure}

Figure \ref{fig:crickets_example} serves as an example of how CRICKETS performed on one of the 32\,MHz subbands. It excelled at marking strong, isolated RFI signals and regions of denser RFI. One of \textsc{CRICKETS} most consistent problems was failing to identify or falsely identifying RFI in the frequency ranges where the sinc-like function of the polyphase filterbank (PFB) response for each 1\,MHz coarse channel approached a local minimum. To greatly improve the reliability of CRICKETS, we are in the process of adding a PFB response subtraction function. A full analysis and plots from each subband are found in the National Science Foundation Research Experiences for Undergraduates (NSF REU) program student report\footnote{\url{https://www.nrao.edu/students/2023/Reports/SofairJared.pdf}}.

From the \textsc{CRICKETS} pipeline, a $CSV$ file and a \textsc{Python} \textsc{Numpy} array were created listing all channels that had excess kurtosis. This information was stored on the COSMIC computers to be applied in the postprocessing steps when evaluating the data collected during VLASS as a first-level removal of significant false positives. We note that not all of the subbands recorded during VLASS by COSMIC were evaluated by \textsc{CRICKETS} due to some data being lost resulting from a storage disk failure. However, in the future, we will work to make this a more robust process, as described in Section 6.

Based on this analysis and plots provided by NRAO for the VLA RFI environment, COSMIC recorded the band between 2.5 and 3.5\,GHz for our observations during VLASS. Therefore, avoid the regions of the spectrum that contain the strongest RFI.

\section{The ARTISTIC Postprocessing Pipeline}
\begin{figure}
    \centering
    \includegraphics[width=0.95\linewidth]{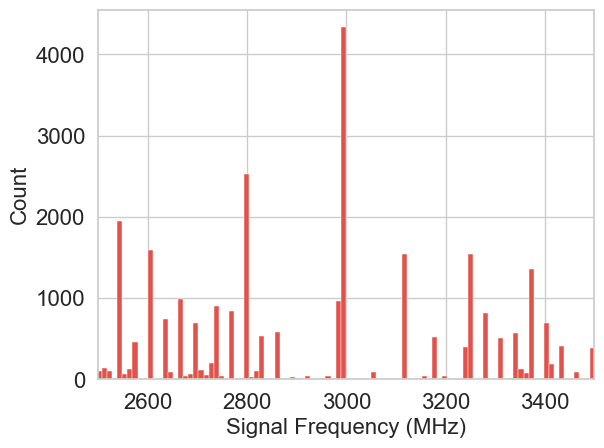}
    \includegraphics[width=0.95\linewidth]{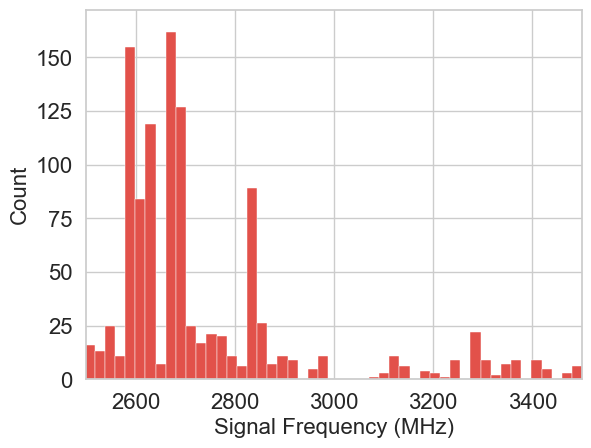}
    \caption{Top: distribution of detected signals by the real-time pipeline before any filtering for RFI. Bottom: number of channels flagged with high kurtosis per frequency bin. When the two plots are compared, there are frequency ranges where there are not many flagged channels but a large number of hits.}
    \label{fig:hits}
\end{figure}

\begin{figure*}
    \centering
    \includegraphics[width=\linewidth]{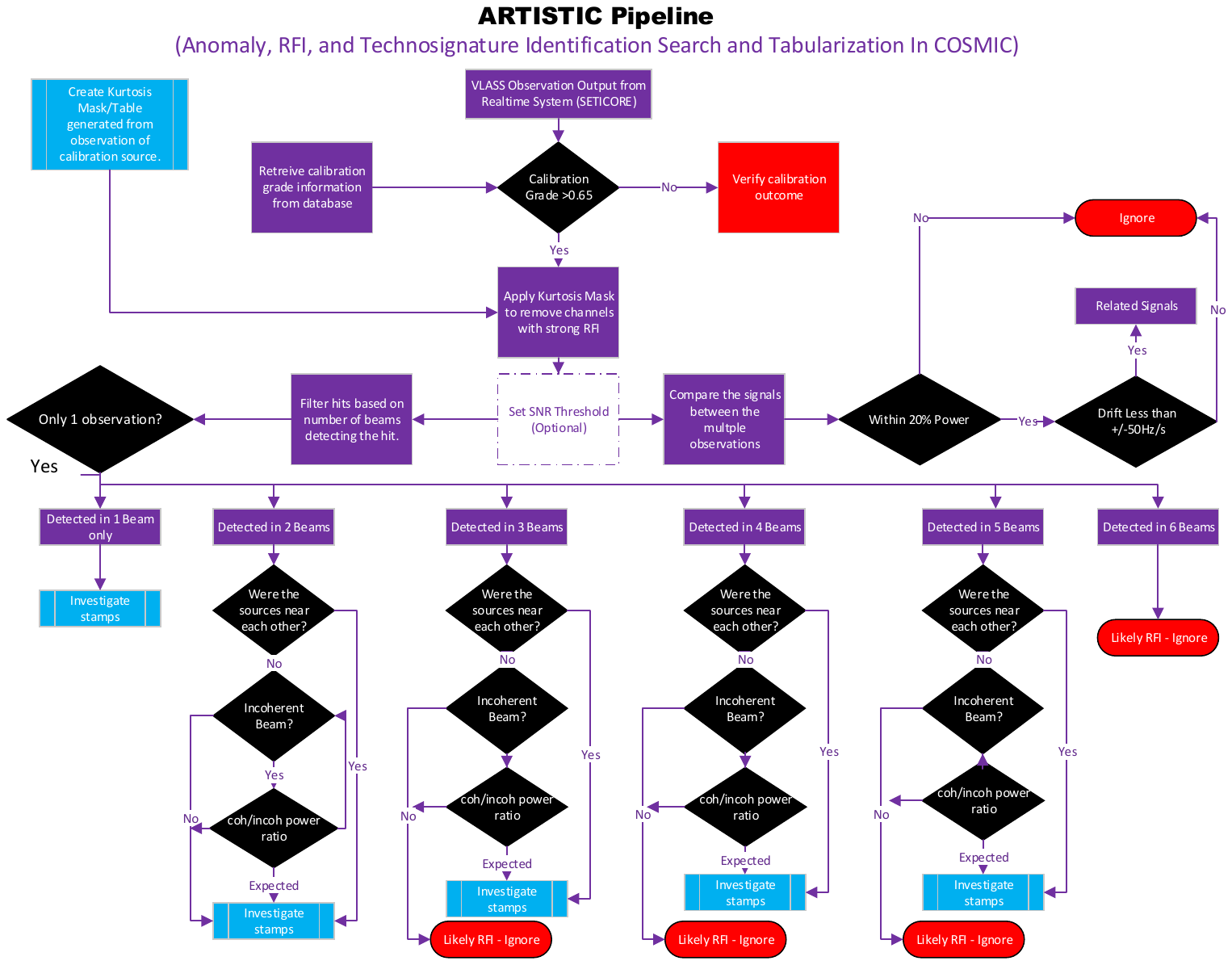}
    \caption{A flow diagram showing the logic behind the ARTISTIC pipeline that can be applied to any set of observations.}
    \label{fig:artistic_single}
\end{figure*}

After passing the data through \textsc{seticore}, potentially millions of hits are identified across the range of frequencies searched, far too many hits to identify likely technosignature candidates individually with visual inspection. To solve this problem, we developed the Anomaly, RFI, and Technosignature Identification Search and Tabularization In COSMIC (ARTISTIC) postprocessing pipeline. In this pipeline, extraterrestrial technosignatures are characterized by a set of assumptions and conditions that, if not met, are used to eliminate hits that do not meet these assumptions. The details of these assumptions are explained below and summarized in Figure \ref{fig:artistic_single}. For the data from 15 April 2023, over 29,000 hits were identified by \textsc{seticore} and distributed across the 2.5--3.5\,GHz frequency band as shown in Figure \ref{fig:hits}, with the majority of false positives associated with RFI. 

Initially, the pipeline evaluates the calibration solutions for the set of observations being evaluated. We first verified the calibration grade score and then reviewed the plots to ensure that the data quality was of scientific merit, as discussed in Section 2.2. If the phase calibration was poor, this would result in coherent beams not being formed at the sources of interest and reduce the overall power of the detected signal. 

Second, the data are evaluated against the textsc{CRICKETS} output described in Section 3.4 and a user-selected SNR threshold (a changeable value determined to best remove additional significant sources of time-variable RFI and spurious antenna artifacts). A description of how the software determines the noise level and the limitations of this method are discussed in detail in \cite{Choza_2023}.  

As a result of using an interferometer for SETI and forming multiple coherent beams, we can filter the results based on this target strategy. There is no localization of a source of emission if it is detected in all coherent and incoherent beams. As a result, these signals are immediately removed from the list of candidate signals. If a hit at a given frequency is only detected within a single coherent beam, we can assume the signal is localized in the sky and thus becomes a potential signal of interest. If the emission is found in multiple beams, then other criteria are considered, such as the ratio of the signal strength in the coherent and incoherent beams and how close the coherent beams are formed. We assume that a signal emitted by a single source should be relatively localized to within the single coherent beam, except when a strong source is present and therefore could have a signal that bleeds into an adjoining beam. As described in Section 3.7.2 in \cite{tremblay_cosmic}, the ratio of the normalized power from the coherent and incoherent beams should be equal to the number of antennas used in the observation.  

The final stage of each track within the ARTISTIC pipeline is to determine if a signal is likely RFI or if it is worth investigating by creating dynamic spectra. At the end of the real-time processing pipeline, a series of files containing segments of raw voltages from each online antenna is saved. We can use these, combined with the visualization tools within \textsc{seticore} and the information about the beam coordinates stored in a beamformer $bfr5$ file, to recreate the coherent and incoherent beams and therefore the dynamic spectra. All signals that were only detected in a single antenna or a small group of antennas were also immediately discounted as being of astronomical origin. This is done by calculating the SNR a the signal position for each antenna, using the \textsc{seticore} viewer software.

Due to the OTF scan strategy, there is an overlap in the sky; therefore some sources are observed multiple times within the same VLA scheduling block. We can use this second observation within a short period of time to apply additional filters, as per Figure \ref{fig:artistic_single}. Any signal of interest that has a time separation of $\sim$10 minutes should be drifted in frequency due to Doppler acceleration but localized to the same position in the sky. We searched for signals that were sky localized and drifted no more than $\pm$50\,Hz\,s$^{-1}$. 

\section{Results}
To evaluate the effectiveness of the ARTISTIC pipeline, we use data from 25 April 2023. This represents a 2-hour observation run with COSMIC, covering 468 beamformed sources and a total of 511 coordinates (when adding in the incoherent beam phase center). 
We found all plots and results from calibration on 15 April 2023 were consistent with high-quality data after the F-engine correction.


The data were compared to the outputs by \textsc{CRICKETS} to look for signals that passed or were contained within the flagged RFI regions. We also use an SNR threshold of 100 to remove signals impacted by intense instrumental artifacts that appear in a single or only a pair of antennas. Upon quality checks of the raw voltages, it was determined that this value removed most artifacts found only a few of the antennas and not suggestive of signals with astronomical origins. With these two filters, we reduced the 29,390 signals down 77\% to 9,708 signals of potential interest.

\begin{figure}
    \centering
    \includegraphics[width=0.95\linewidth]{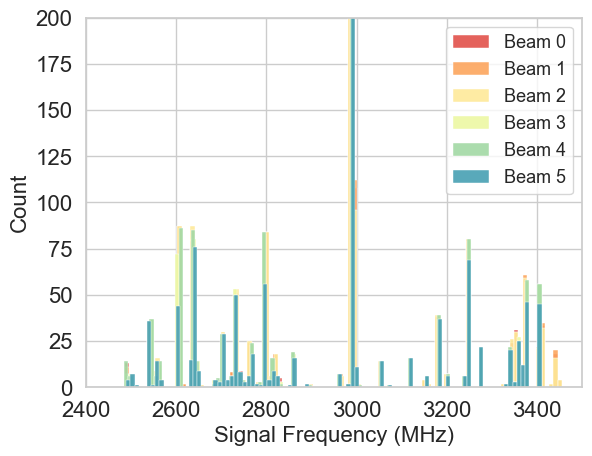}
    \caption{A histogram demonstrating the signals detected for each frequency per each beam. This shows that many of the signals are in similar frequency ranges and at a similar number of detected signals per frequency.}
    \label{split}
\end{figure}

As shown in Figure \ref{split}, the majority of the signals detected at each frequency are in all of the beams. After the first two stages, we have
\begin{itemize}
    \item 1646 in Beam 0;
    \item 1650 in Beam 1;
    \item 1657 in Beam 2;
    \item 1615 in Beam 3;
    \item 1625 in Beam 4;
    \item 1515 in Beam 5 (incoherent beam);
\end{itemize}

\begin{figure*}
    \centering
    \includegraphics[width=\linewidth]{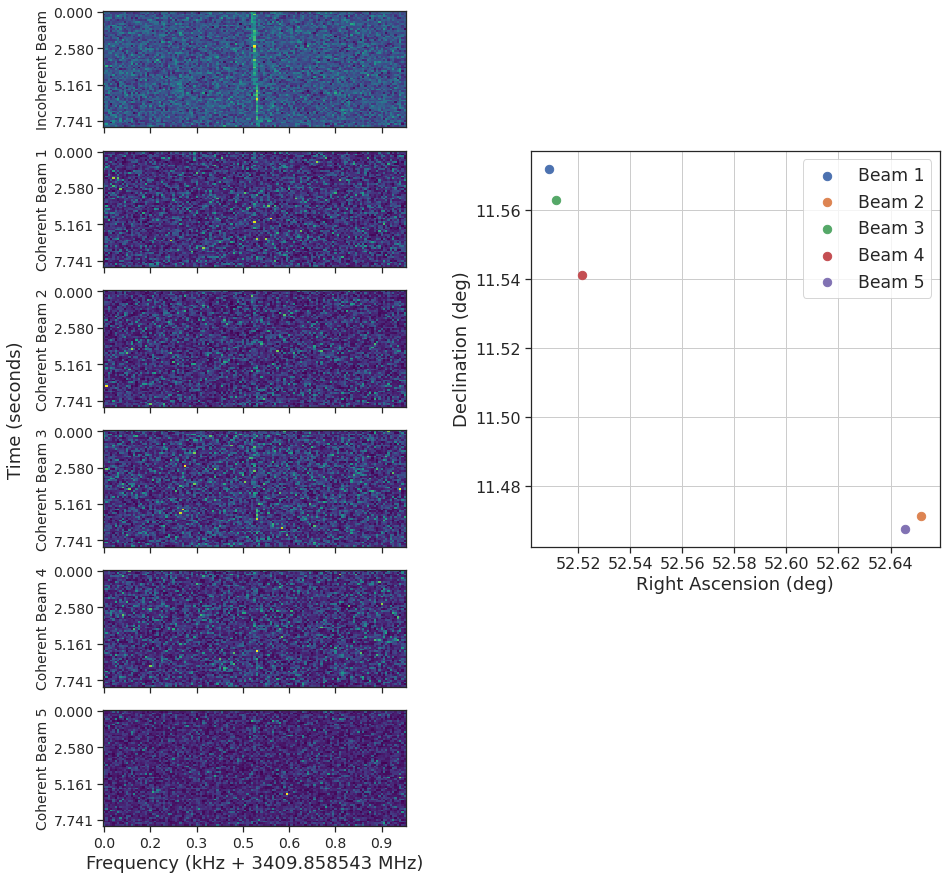}
    \caption{An example of the dynamic spectra from a single field in which the incoherent sum contained the main signal but the coherent beams each have a faint trace of the signal, indicating it is not likely an astronomical source of emission. All sources chosen are within the FWHM of the primary beam.}
    \label{Waterfall}
\end{figure*}

When filtered, no signals were contained in only a single (coherent or incoherent) beam. An example of one of the signals detected by COSMIC is shown in Figure \ref{Waterfall}, where the signal is seen in multiple coherent beams and the incoherent beam.

To determine how bright a signal would need to be for each source in order for us to detect it, we compute the minimum equivalent isotropic power (EIRP$_\mathrm{{min}}$) for the 511 sources,

\begin{equation}
    EIRP_\mathrm{{min}} = 4 \pi d^2 F_\mathrm{{min}}
\end{equation}

where the relationship to the detectable power is based on a distance squared. The $F_\mathrm{{min}}$ value is determined by dividing the minimum flux density the bandwidth of the transmitting signal. For this experiment, we use a minimum flux density of 13.92\,Jy which is the 8$\sigma$ value of the flux density limit from the VLA sensitivity calculator. The bandwidth is set as equal to the channel width of $\sim$8\,Hz. 

\begin{figure}
    \centering
    \includegraphics[width=0.95\linewidth]{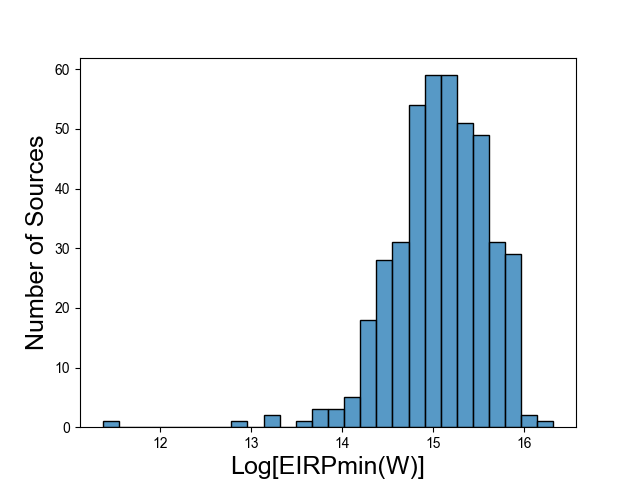}
    \caption{A plot showing the distribution of EIRP values calculated for each of the 511 sources evaluated during this work. The values range from 10$^{11}$ to 10$^{16}$\,W, providing some of the lowest values for technosignatures to date.}
    \label{EIRP}
\end{figure}

As shown in Figure \ref{EIRP}, the EIRP$_\mathrm{{min}}$ values range from 2.32$\times$10$^{11}$ to 2.09$\times$10$^{16}$\,W. Based on the distances calculated using parallax in \cite{Czech_2021}, the sources range in distance between 4.3 and 1296\,pc. By using the value for the Arecibo radio telescope transmitter power (L$_A$) of 2$\times$10$^{13}$\,W from \cite{Siemion_2013}, we can calculate the ratio (EIRP$_\mathrm{{min}} \div$L$_{A}$) of this value for each source in our survey. As a result, we established a value range between 0.012 and 1045, which suggests that if an Arecibo transmitter was located on a planet orbiting the nearest star 4.3\,pc away, we would be able to detect it at least at an 8$\sigma$ level.

\section{Discussion}
The field of astronomy is experiencing a relatively new problem called "big data." The idea is that we are receiving increasing quantities of data that must be sorted in new ways in order to find information of scientific interest. Although using the world's population and their excitement to filter out signals of interest through their collective enthusiasm is a feasible method for certain kinds of projects \citep{Li_2024}, we could quickly overload people if we did this with all of the data generated by COSMIC. Therefore, creating postprocessing pipelines through a series of logical steps that can be followed by humans or through machine learning/neural networks is necessary as we progress with our data collection.

\begin{figure}
    \centering
    \includegraphics[width=0.95\linewidth]{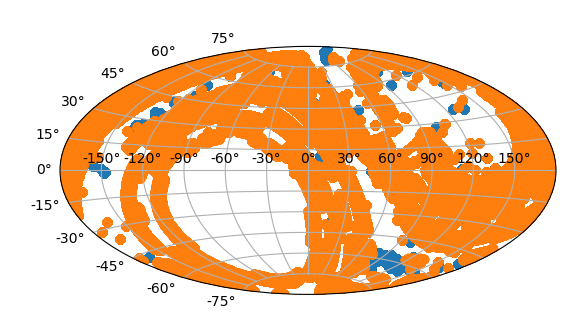}
    \caption{A plot in galactic coordinates of all the coordinates currently in the database observed from 29 March 2023 to 14 July 2024. The orange points represent data from frequencies below 4\,GHz and the blue points are from data collected above 4\,GHz. }
    \label{Coordinates}
\end{figure}

As shown in Figure \ref{Coordinates}, within the last 11 months of operation, COSMIC has observed over 950,000 fields and is rapidly becoming one of the largest SETI experiments ever designed. If any signal is identified by the postprocessing pipeline described in this paper or another, we have successful proposals with the VLA, the Very Long Baseline Array, the Allen Telescope Array, and the Parkes 64m Telescope to follow up on those signals within a 1-week time frame. However, we note that at this time, the time frame for detecting signals for follow-up will take months until postprocessing becomes more automated. However, this post detection protocol will allow us to identify any potential signals of interest by using different signal pathways, telescopes, and locations. This network is critical to understanding the nature of any signal detected that cannot immediately be dismissed as RFI.

\section{Conclusion}
In this paper, we describe COSMIC's operation relating to the observational strategy during the NRAO-hosted VLA all-sky survey, which observes the entire sky at declination's above --40 degrees. The data are processed through an autonomous pipeline that generates a set of potential signals and logs the metadata into a MySQL database for scientists to filter through. 

Using a test field of 30 minutes of data consisting of 511 sources, we describe a potential postprocessing pipeline that uses a series of filters to separate RFI from signals with an astronomical origin. In this search, no potential technosignatures were identified. In the future, a faster and more automated version of this pipeline would allow for fast identification of potential candidates, which is a goal the team is striving toward.  

This work overall represents an important milestone in our search. With the rapidly growing database, we need new methods for sorting through the data, and this paper describes a rapid and viable filtering mechanism.

\paragraph{Data Availability Statement}
All software used in this publication is available on GitHub and is referenced throughout the paper. The observations using the COSMIC system on the VLA were collected during commensal observations and can be accessed by contacting the lead author on this team. If particular data or computing resources are needed, they can be provided by direct inquiry to the Breakthrough Listen team.

\paragraph{Software}
\begin{itemize}
\item {\sc topcat} -- \cite{Topcat}
\item {\sc CASA} -- \cite{CASA}
\item{NumPy v1.11.3 \citep{NumPy}, AstroPy \citep{Astropy}, SciPy \citep{SciPy}, Matplotlib \citep{Matplotlib}}, Pandas (\url{https://pandas.pydata.org/docs/})
\item {\sc pyUVdata} -- \cite{pyUVdata}
\item {\sc CARTA} -- \cite{angus_comrie_2020_3746095}
\item {\sc Blimpy} -- \url{https://github.com/zoips/blimpy}
\item {\sc ansible playbook} -- \url{https://docs.ansible.com/}
\item {\sc BLADE} --\url{https://github.com/luigifcruz/blade}
\item {\sc seticore}  -- \url{https://github.com/lacker/seticore}
\item {\sc hashpipe} -- \cite{MacMahon_2018}
\item xGPU -- \url{https://github.com/GPU-correlators/xGPU}
\item COSMIC Software -- \url{https://github.com/COSMIC-SETI}
\item target selector -- \url{https://github.com/danielczech/targets-minimal}
\item {\sc Redis} --\url{https://redis.com/}
\end{itemize}

\begin{acknowledgments}
We gratefully acknowledge the foundational support from John and Carol Giannandrea that has made COSMIC possible. We acknowledge additional support from other donors, including the Breakthrough Prize Foundation under the auspices of Breakthrough Listen. The National Radio Astronomy Observatory is a facility of the National Science Foundation operated under a cooperative agreement with Associated Universities, Inc. LS was funded as a participant in the Berkeley SETI Research Center Research Experience for Undergraduates Site, supported by the National Science Foundation under Grant No. 2244242. JS was funded by the NSF through the REU program at NRAO.
\end{acknowledgments}

\bibliography{cosmic}{}
\bibliographystyle{aasjournal}



\end{document}